# A comparative study of approaches in user-centred health information retrieval

Harsh Thakkar[a], Ganesh Iyer[b], Prasenjit Majumder[c]

*Dhirubhai Ambani Institute of Information and Communication Technology, Gandhinagar, Gujarat, India-385007.*

*{[a]harsh9t,[c]prasenjit.majumder}@gmail.com, [b]me@ganeshiyer.net*

**Abstract**

In this paper, we survey various user-centred or context-based biomedical health information retrieval systems. We present and discuss the performance of systems submitted in CLEF eHealth 2014 Task 3 for this purpose. We classify and focus on comparing the two most prevalent retrieval models in biomedical information retrieval namely: Language Model (LM) and Vector Space Model (VSM). We also report on the effectiveness of using external medical resources and ontologies like MeSH, Metamap, UMLS, etc. We observed that the LM based retrieval systems outperform VSM based systems on various fronts. From the results we conclude that the state-of-art system scores for MAP was 0.4146, P@10 was 0.7560 and NDCG@10 was 0.7445, respectively. All of these score were reported by systems built on language modelling approaches.

*Keywords:* concept-based information retrieval, clinical document retrieval, query expansion, language models, vector space models.

## 1. Introduction

Clinical health information retrieval has become a necessity today due to the enormous growth in health related information over the internet. It has become extremely difficult to cope up with the pace of release of new research in the biomedical domain. With the spreading awareness, searching online health related web-forums and other sources has become a common habit. A recent survey [1] suggests that around *eighty percent* population of the U.S.'s search engine users seek information on specific diseases or disorders.

Mining health related information from a variety of biomedical data (viz. prescriptions, clinical reports, blogs, forums, etc.) is a domain-specific information retrieval task. The beneficiary class of this task comprises of a vast community of medical practitioners, clinical attendants, patients (and their next-to-kins), researchers and anyone with precise health information related needs.

Thus, it has become inevitable to design user centred health information retrieval systems that cater the users with context-based specific information as desired in real time. Taking this challenge as an opportunity communities like TREC[*] (Text REtrieval Conference), CLEF[†] (Conference and Labs of the Evaluation Forum) have undertook the task of harvesting domain-specific resources, organizing worldwide challenges/competitions, and encourage research in the field of targeted information retrieval. One of such a challenge is CLEF ehealth task. In this paper, we present a survey of various techniques applied for targeted health information retrieval.

This paper is organised as follows:

Section 2 presents the other similar organizations dedicated to health IR. Section 3 narrates the aim of the task, data and subsequent evaluation parameters in brief. Section 4 presents a detailed survey of various methodologies and techniques applied and its underlying analytical critic. Section 5 concludes the paper with the authors findings and comments.

## 2. Related work

User centred information relates to individual patients with a goal to present health professionals about the condition of a patient. It classically involves the patient's medical history, current diagnostics and/or prescribed drugs record based on the specific format used. These records generally contain unstructured data or semi-structured data, due to a mixture of computer generated results and free/narrative text (e.g. medical attendants daily notes, radiology report, etc).

The previous research works have primarily investigated the query logs from commercial search engines [2]. There is no system at present that contemplates the information needs of the user in his/her context and returns relevant documents, to the best of our knowledge. Thus, there is a lack of attention to the development of this kind of evaluation campaigns addressing the issues of information needs of a layman patient or the next-of-kin. Their queries and searching time have reported to tend much shorter compared to those considered in earlier health information retrieval standards [3, 4].

---

[*] http://trec.nist.gov/
[†] http://www.clef-initiative.eu/

The medical database PUBMED[*] (also known as MEDLINE contains over 19 million references to medical documents. Annually, 500,000 new citations are added on an average in this database. OHSUMED, which was published in 1994, is the first collection consisting of medical data used for IR evaluation [5]. This collection contained an approximate of three fifty thousand abstracts extracted from medical journals on the MEDLINE over a period of five years (i.e. from 1987–1991). This collection was first created for TREC 2000 Filtering Track, but later was also used for a variety of research on health information retrieval [22].

## 3. The Task

The task was to develop an efficient user-centred[20] or context-based[21] health information retrieval system which returns relevant documents to users/patients in the context of their query, to satisfy their health-related information needs.

### 3.1. Corpus

The corpus was obtained from the ShARe/CLEF community. This corpus consists of a web crawl of large number of medical documents from the Khresmoi[†] project The dataset comprises of a 1.6 million English documents covering a wide range of medical subjects. This collection is a result of the efforts of a wide range of organizations including Health on the Net Foundation[‡] certified website (e.g. Genetics Home Reference, ClinicalTrial.gov, Diagnosia[§]).

The size of the corpus is 6.3 GB compressed and about 43.6 GB uncompressed.. This document collection of CLEF eHealth 2014, is a group of .dat files. Each of the .dat files in this collection consists one particular topic or disease or medical issues.

The format of the data in the .dat files is described below:
- (#UID): a unique identifier for a web page
- (#DATE): the date of crawl (YYYYMM),
- (#URL): the link to the original web page, and
- (#CONTENT): the raw HTML content of the web page

### 3.2. Discharge summaries

---

[*] MEDLINE is the U.S. national library of medicine's (NLM) premier bibliographic database of journal life sciences articles with a focus on biomedicine. Online at http://www.nlm.nih.gov/bsd/pmresources.html
[†] EU-FP7 Khresmoi (http://khresmoi.eu/) project
[‡] http://www.healthonnet.org
[§] http://www.diagnosia.com/

The discharge summaries are the medical description of patient's conditions at the time of discharging from the hospitals. It contains vital information about the medical history, disease diagnosed, treatments given, drugs prescribed and etc. about the patient. These discharge summaries are created from the de-identified MIMIC-II[**] database (Multi-parameter Intelligent Monitoring Intensive Care). Non-disclosure agreement and privacy contract were to be signed and submitted in order to obtain the access of this dataset.

### 3.3. Evaluation metrics

The retrieval systems were evaluated based on various parameters which are: MAP (Mean Average Precision), P@10 (Precision at 10 documents), and NDCG@10 [Normalised Discounted Cumulative Gain at 10 documents], with P@10 and NDCG@10 being the primary and secondary evaluation measures.

## 4. Comparison of approaches and Discussion

In this section we present and discuss rsults of various approaches deployed to address the problem of user-centred I.R. We broadly classify the approaches used under two classes, i.e. approaches using *language modelling models* and approaches *using vector space models*. A gist of the approaches and their relative performance can be obtained from table 1, later in this section.

**LM based approaches:**
A language model basically is a statistical distribution model that assigns probabilities to a sequence of terms, which depict the likelihood of their occurrences in the text. Language model based systems work on probabilities for each term encountered and these probabilities and independent of the nature of document.

In their work, [6] made used the Indri search engine platform as their baseline. They used language modelling approach integrating Dirichlet smoothing. UMLS and Metamap were incorporated for the purpose of query expansion. In addition, query expansion was also performed using mutual information from the user query file and an intersection of common words from query file and medical thesaurus was used for concept-based information retrieval. Efficacy of their system was reported to be the highest (in P@10 value) yielding a score of 0.4016 MAP and a P@10 value of 0.7560.

In their work [7] the authors propose a multiple-stage re-ranking method. Lucene is used as the search engine for baseline. Like [6] they also use Dirichlet smoothing. They incorporate query expansion using abbreviation and terms extracted from discharge summaries. They use a combination of methods for scoring/ranking documents, which include centrality-based scoring by means of implicit links between documents, and pseudo relevance feedback.

---

[**] MIMIC-II databse online at https://mimic.physionet.org/

Their system reported score of 0.3989 MAP and a P@10 value of 0.74.

In a another variant approach of language modelling [8] made use of indri search engine platform as their baseline system. The approach differs in choice of term synonyms as they incorporated the use of morpho-syntactic variants. These are the terms that derive from the same root. The authors considered the roots rather than different variants of words to extract terms from the UMLS. Nonetheless, the performance of the system was reported to be inconsistent, reason for the same are unclear. Their system reported a score of 0.4021 MAP and a P@10 value of 0.67.

Yet another approach by [9] made use of Indri as their baseline system for indexing the document collection. They incorporate query-likelihood model along-with pseudo relevance feedback for query expansion. They made use of **MeSH**[*] and discharge summaries to extract terms as matching concepts for query expansion. Their system reported score a score of 0.4016 MAP, yielding the highest MAP score amongst the compared approaches and a P@10 value of 0.7060

In a similar approach as [6,7]; [10] also used indri as their baseline system along-with Dirichlet smoothing language model respectively. They also incorporated query expansion using the Metamap thesaurus, with extracted terms from original query matched with those from discharge summaries. They also integrated learning to rank methods to experiment with the "quality feature". This features counts the frequency of terms prior-hand and checks for which of them appear in the documents. This system reported score of 0.3494 MAP, and second highest P@10 value of 0.75.

In their work [11], performed a sentence level retrieval on the cleaned dataset. They considered Indri as their baseline engine. For query expansion, they incorporated pseudo relevance feedback with the description and narrative content from the discharge summary files. In a parallel setup they used modified markov random field model (MRF) for expanding queries using terms extracted from the medical phrases from Metamap. An improvement of eight to fifteen percent was reported by the authors as compared to their earlier setup. Their system reported a score of 0.3589 MAP and a P@10 value of 0.69.

An interesting variant was presented by [12] by using Hiemstra language model[18] along-with terrier as their baseline system. Query expansion was performed using pseudo random feedback using Medline resource which reported to have improved performance of the system. They incorporate the HTMLstrip[†] and Boilerpipe[‡] resources to reduce the dataset size to six percent of the original, eliminating the non-relevance terms (e.g. html tags, advertisements, header-footers, etc.). HTMLstrip was reported to be the best approach when compared to boilerpipe as it only removed the html tags which were non-relevant, while important terms were found in header-footer section. These terms were reported to have a high impact on the over-all system performance. The system reported a score of 0.1677 MAP and a P@10 value of 0.5360.

**Vector Space model (VSM) based approaches:**

The Vector space models are one of the most used models, in information retrieval domain. Vector space model represents pieces of text(any object) as vectors of identifiers such as terms in the text. It was first implemented in the **SMART information retrieval system**[19]. Since then many variants have been proposed, generalised vsm, weighted vsm (TF-IDF weights), etc. The advantages of vsm based systems include: ease of implementation (as vsm is based on linear algebra), it computes the degree of similarity between queries and pool documents, it also supports partial matching. However, it shows very poor similarity values when representing long text documents. Also words that appear with very less frequency but are highly relevant to the current context of the query are neglected resulting in poor performance of the overall system. In this section we review some of the modern vsm based approaches adapted by researchers.

In their work [16] the authors have used the vsm model for purpose of user-centric information retrieval as it allows weighted vectors to represent the documents and not binary. The degree of similarity was measured by the cosine degree (cosine of the angles) between the document vector and the query vector. They calculate the degree by multiplying the weighted TF and IDF measures to represent the connection between a query term and the total pool of documents. The Terrier retrieval system, with default setup is used for stopword elimination, tokenization, and stemming. Their system obtained a score of 0.167 (MAP) and a P@10 value of 0.55.

Likewise, in an another vsm variant by [13] used the Lucene[§] as their baseline system for tokenization, stemming, and etc. They incorporate query expansion by using a pseudo-relevance feedback method. In order to perform the expansion of the user query, they append words extracted from Medline biomedical dictionary. These terms are extracted based on the Rocchio's formula, which was used for stemming in the SMART system, with custom value of Pseudo-relevance feedback. Their system reported score of 0.20 MAP and a P@10 value of 0.5540. A minor improvement in the MAP and P@10 values is observed as compared to [16].

---

[*] MeSH (Medical Suject Headings) is the compendium of controlled medical vocabulary for indexing articles for PubMed. It is maintained by the National Library of medicine library, Online at
http://www.ncbi.nlm.nih.gov/mesh

[†] HTMLstrip removes the html tags from the text documents. Online at
http://search.cpan.org/~kilinrax/HTML-Strip-1.10/Strip.pm

[‡] Boilerpipe removes the surplus clutter from the text. Online at https://code.google.com/p/boilerpipe/

[§] Apache lucene is a open source information retrieval search engine library written originally in java by Doug Cutting. Online at http://lucene.apache.org/core/

In an another system, [14] used terrier as their baseline engine. Unlike [13], they incorporate prediction method (based on probabilistic naïve bays classifier) to assure whether there is a necessity for query expansion for that particular query. If the results are in favour, the query is expanded, otherwise not. The features are extracted by the classifier prior hand, by training on the document collection (dataset). This system reported a score of 0.305 MAP, and a P@10 value of 0.67. It can be observed that query prediction improves the previous vsm based system performances by am approximate factor of 12% and 10% in terms of P@10 and MAP respectively.

In a variant reported by [15], lucene was used as baseline engine, which indexed both unigrams and bigrams respectively. Query title (from the user query file) along with synonym concept (terms) extracted from MeSH, was solemnly used for query query expansion. Only a specific number of terms (i.e. only certain medical terms and not its sub-parts) were selected for query expansion. This system reports a best performance of 0.2315 MAP and a P@10 value of 0.5460. From the results it can be concluded that only the query title might not be sufficient for query expansion as terms present in the description of the query also put emphasis on the nature of the search.

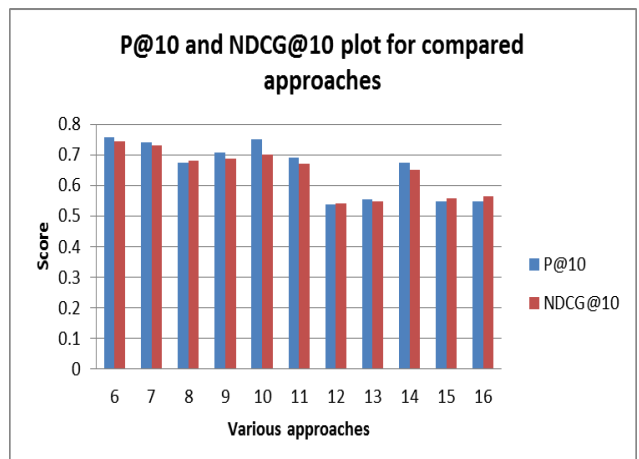

**Figure 1:** Performance comparison of various approaches based on P@10 and NDCG@10 values.

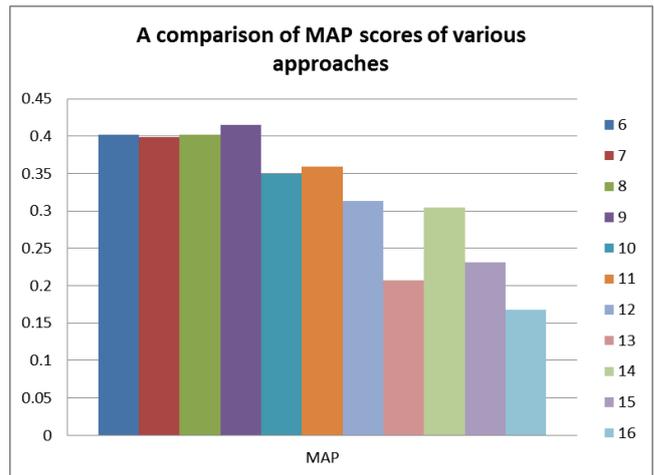

**Figure 2:** Performance comparison of various approaches based on MAP values.

## 5. Author remarks and Conclusion

The state-of-art results are by works of [6], with a **P@10 of 0.7560** and a **NDCG@10 of 0.7445** and [9] reported the highest value of **MAP at 0.4146** (see fig. 1 and fig. 2). The work of these authors are based on language modelling retrieval methods, perform query expansion and two of them use Metamap and UMLS (medical thesaurus). It is clear from results that language modelling methods report the state-of-art results in the domain of biomedical document retrieval, beating the vsm based systems by a factor of 3%-8% in terms of P@10 and almost by 20% in MAP scores. Furthermore, approaches integrating variants of query prediction and learning to rank methods are being developed with a high promise of delivering superior performance in the biomedical information retrieval domain in near future.

Table 1
Comparison of system performances of language model and vector space model approaches. The highlighted figures denote the highest scores in each parameter respectively.

| Work | Approach | Query expansion | Other resources used | System performance | | |
|---|---|---|---|---|---|---|
| | | | | P@10 | NDCG @10 | MAP |
| [6] | Language model | Metamap, UMLS, Mutual information | Metamap, UMLS | **0.7560** | **0.7445** | 0.4016 |
| [7] | Language model | Abrv. + Pseudo random feedback | None | 0.74 | 0.73 | 0.3989 |
| [8] | Language model | UMLS synonyms, abbreviations, and FASTR morpho-syntactic variants | TreeTagger, FASTR, Ogmios NLP, YATEA, UMLS | 0.6740 | 0.6793 | 0.4021 |
| [9] | Language model | Query-likelihood + pseudo relevance feedback | MeSH, Metamap | 0.7060 | 0.6869 | **0.4146** |
| [10] | Language model | Intersection of terms from query and discharge summaries | Metamap, UMLS | 0.75 | 0.70 | 0.3494 |
| [11] | Language model | Markov random field for medical terms + pseudo relevance feedback | Metamap | 0.69 | 0.6705 | 0.3589 |
| [12] | Language model | Pseudo relevance feedback | BoilerPipe, JusText, HTMLstrip | 0.5360 | 0.5408 | 0.3134 |
| [13] | Vector space model | Pseudo relevance feedback | Medline | 0.5540 | 0.5471 | 0.2076 |
| [14] | Vector space model | Naïve bayes probabilistic expansion | Weka | 0.6740 | 0.6518 | 0.3049 |
| [15] | Vector space model | UMLS Synonyms | Metamap, UMLS, MeSH | 0.5460 | 0.5574 | 0.2315 |
| [16] | Vector space model | Weighted vectors for query terms | none | 0.5460 | 0.5625 | 0.1677 |